# The 1908 Tunguska event and some distant phenomena

Andrei Ol'khovatov

https://orcid.org/0000-0002-6043-9205

Independent researcher
(Retired physicist)

Russia, Moscow
email: olkhov@mail.ru

submitted to arxiv.org

**Dedicated to the blessed memory of my grandmother ( Tuzlukova Anna Ivanovna ) and my mother ( Ol'khovatova Olga Leonidovna )**

**Abstract.** This paper is a continuation of a series of works, devoted to various aspects of the 1908 Tunguska event. This usually refers to an explosive phenomenon associated with the appearance of a forestfall, named nowadays as the Kulikovskii one. However, several other notable natural phenomena occurred in the Central Siberia on June 30, 1908. This paper considers geomorphological features, reports of substance discoveries, forestfalls, and earthquakes. The general conclusion is that the Tunguska event was a very complex phenomenon. Research of the Tunguska event requires the participation of experts in various fields.

## 1. Introduction

This paper is a continuation of a series of works in English, devoted to various aspects of the 1908 Tunguska event [Ol'khovatov, 2003; 2020a; 2020b; 2021; 2022; 2023a; 2023b; 2025a; 2025b; 2025c; 2025d; 2025e; 2025f; 2025g; 2025h]. The works can help researchers to verify the consistency of the various Tunguska interpretations with actual data. A large number of hypotheses about its causes have already been put forward. However, so far none of them has received convincing evidence. As it is written on the title web-page of the web-site created by KSE (see below about KSE) tunguska.tsc.ru/ru/ (translated by A.O.):

"About a thousand researchers have devoted years of their lives to the Tunguska phenomenon. However, there is still no well-founded scientific understanding of what happened over the Siberian taiga on June 30, 1908."

This is probably why new hypotheses appear almost every year, not only in the mass-media, but also in scientific literature. At the same time, any hypothesis should not contradict the known facts about the event. Unfortunately, the authors of new hypotheses, as well as the authors of popular science articles, often use data, many of which turned out to be not entirely accurate, or even incorrect. The author of this paper hopes that it will help both the authors of various hypotheses and their readers to evaluate the validity of the proposed hypotheses.

In this paper some phenomena are considered which are outside the area of the forest-fall in general.

Let's start with brief info about research of the Tunguska event. The Committee on Meteorites of the USSR Academy of Sciences (KMET) stopped research the area of the Tunguska event in the early 1960s. Later amateurs (consisting mainly of scientists, engineers and students) most of whom united under the name Kompleksnaya Samodeyatel'naya Ekspeditsiya (KSE) continued research (KSE started research in 1959). Since the late 1980s foreign scientists take part too.

Please pay attention that so called the epicenter of the Tunguska forestfall (the forestfall is named "Kulikovskii") is assigned to 60°53' N, 101°54' E, and in this paper is called the epicenter.

In this paper its author (the author of this paper i.e. A.O.) for brevity will be named as "the Author". The surname Vasil'ev can also be translated as Vasilyev in some references.

## 2. Some distant phenomena

**Geomorphological features.** Let's start with the V. Bronshten story about Konstantin Suvorov expedition to Tunguska in 1934. Bronshten wrote [Bronshten, 1997], translated by A.O.:

"Suvorov also visited the so-called "dry river" in the area of the Verkhny Lakur River, which was described to Kulik by the Evenks. Ivan Peskov told him that a thunderstorm and a storm had torn up the forest, and an explosion had created a trench with a crater on the surface of the earth. There was also the Evenk Onkoul's lodge/labaz nearby, which had burned down in 1908. The distance to this location was two days' walk. Together with Peskov and following his instructions, Suvorov conducted a successful reconnaissance tour. At a distance of 43 km from the Stojkovic's mountain in the south-southwest direction (azimuth 225), they found a broken-line ditch. The sides of the ditch had eroded and leveled out. The fractures and steep slopes were smoothed out and overgrown with shrubs. Instead of a crater, the trench ended in a saucer-like depression with a diameter of 7-8 meters. According to Suvorov, this was a fault line that had been exposed to the surface, possibly caused by an underground explosion."

To the Author this feature resembles the feature left by a ball-lightning event occurred in County Donegal, Ireland in 1868 [Vandevender et al., 2008].

There is another interesting structure, but with unknown dating. Here is an abstract of [Dorofeev, 2008]:

"About an unusual explosive structure near the Tunguska tree fall

V.A. Dorofeev
FSUGE "Urangeo", Moscow, Russia

From 1976 to 1979, several geologic parties searched for Iceland spar in the country between two rivers the Chunya and the Chamba, including two sites: in the middle coarse of the Nizhnyaa Dulushma stream and, 30–40 km to the north, at a narrow watershed of the Kimchu, Nizhnyaa and Verhnyaa Dulushma rivers. One of my reconnaissance routes passed along a left side of a brook valley with latitudinal stretch, a left confluent of the Yuzhnaya Chunya river (several tens of kilometers to the northeast form the epicenter of the Tunguska explosion). The route run through the field of development of grey-colorful loosely consolidated silt-psammitic tuffites of early Triassic age, very typical for the upper part of a stratigraphic section of the Korvunchana suite. An isoline of relief on the other side of the valley made a distinctive bend at one of the sections of my route. Such situation in locality must be connected with a terrace. But river terraces for upper reaches of small streams are extremely atypical. One could most likely expect to meet there an outcrop of a bedded body of magmatic rocks – dolerite or metasomatic rocks with quartz-carbonate composition. However I found a turfy terrace made from a breccia – a result of a powerful explosion. The breccia consisted of debris of basaltic lava from 1 to 10–15 cm in size. I was astonished by a variety of colors from grey-green to violet-brown, with predominance of various hues of red. The debris were consolidated by a chemical cement – zeolite (heulandite), and this was also unusual (because analcime is typical for cements of tuffs at that location). The breccia represented a dense and strong rock, which, judging by the terrace, lied subhorizontally. Father, the route run in the field of development of these breccias consolidating a gently sloping watershed. At a distance of 2–3 km I found a hole similar to a crater which had one wall destroyed by erosion, so that in view from above the structure had a shape of a sector covering 300–330 degrees. Its upper diameter was 100–150 m, the diameter at the bottom was about 50 m, and the depth was about some tens of meters. I saw the same breccia at the slopes and bottom of the crater but the debris reached 1–1.5 m in size there. A spring welled out at the bottom, and a streamlet flowed out through a narrow mouth of the destroyed wall and fell into the main valley. Along the periphery of the structure, I found several outcrops of quartz-carbonate metasomatites, which were bow-shaped in view from above, with a curvature center coinciding with the bottom of the hole. Later, during office studies, analyzing geophysical data, I found that the field of the breccia spreading is associated with a magnetic anomaly of high intensity. The information about this finding is in the Report of the geological party on the works of 1978–1979, which is kept in the archives of expedition "Spar" and in Federal funds. (1) I could not determine, even roughly, the area of breccia spreading, however, I think that it is longer than 3 km in linear dimension. A lower boundary is early Triassic, but the upper one is not determined because there are no dated sediments which would intersect the breccia, and the thickness of modern (Holocene) sediments is minor. (2) The explosion, to all appearances, was of near-surface character, because rocks of pre-volcanic substrate are absent in breccias structure, as well as tuffites embracing them. (3) A clear negative shape of relief was evident. There is a rule in geomorphology: if a structure is older, it is worse expressed in relief, and vise versa. Therefore, the suggestion about a relatively young age of the structure does not contradict the actual situation. (4) Connection between the Tunguska "meteorite" and the explosive structure is unclear, however, it seems quite possible because the finding is extremely uncommon."

Interesting iron formations were discovered under the surface lay of soil in 2013 [Dmitriev, 2016] at the distance ~40 km and azimuth ~80º from the epicenter. 12 samples of the iron were collected, total weight 96 kilograms - see Fig.1 from [Dmitriev, 2016].

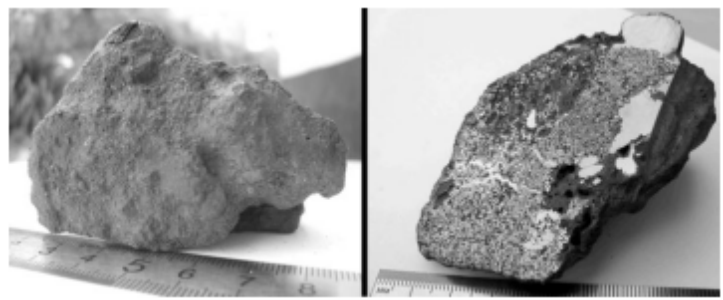

**Fig.1**

One sample weighing 3.5 kg was different from the rest, an alloy of iron and dark green glass - see Fig.2 from [Dmitriev, 2016].

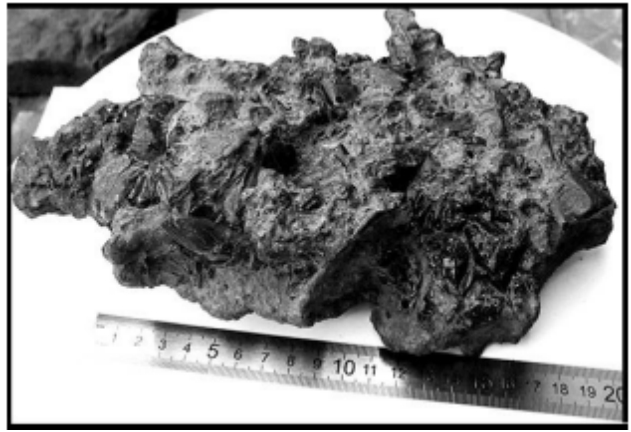

**Fig.2**

Figure 3 (from [Dmitriev, 2016]) shows an enlarged fragment of the central part of Figure 2.

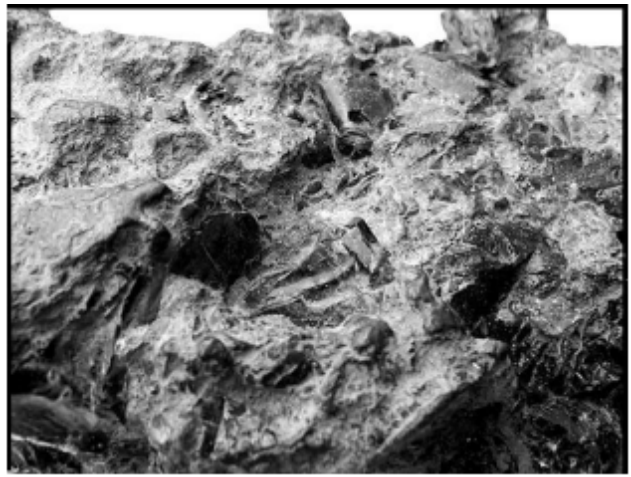

**Fig.3**

Some samples are a calcium aluminosilicate - see Fig.4 from [Dmitriev, 2016].

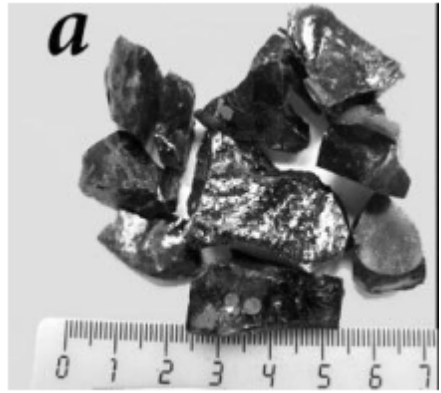

**Fig.4**

In addition, small pieces of high-sodium glass were found - see Fig.5 from [Dmitriev, 2016].

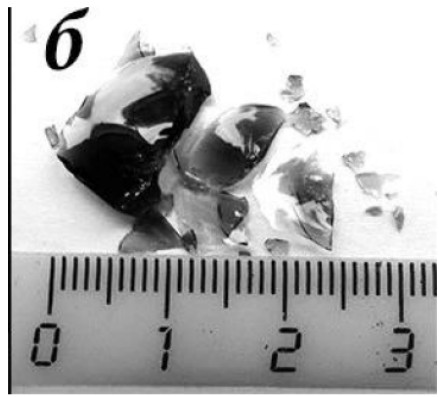

**Fig.5**

The dimensions of the investigated area: 600 m x 2200 m, The samples were at the depth 15 - 45 cm.

The weight of the discovered iron samples is from 1 kg to 42 kg. Samples have splinter-torn structure. The shape is convex like a broken lens from glasses. Sample thickness is from 5 cm to 18 cm. All samples have melting crust. In places of splitting the melting crust is much thinner. Most of the samples consist of native iron with the inclusion various minerals. The iron content is 40-60%.

Completely amorphous small glasses up to 1.2 cm in various shapes, uneven stripes and droplets, also fragments of basalt structure and fine pumice were discovered on the surface of some samples. The samples are not melted in the central internal part. Their mineral composition: metallic native iron, magnetite, quartz, orthoclase, albite, wollastonite.

As for the single sample of iron+glass (Fig. 2); glass – high-sodium silicate, well-melted, amorphous 92 – 100%, formation temperature 2000° - 2550°. Small and large fragments of glass are enclosed in native metallic iron. Native iron in the form of balls is present inside the glass and has a purity of 97- 100%. Small pieces of high-sodium glass (Fig. 5) were examined using a microprobe. These glasses and calcium aluminosilicates contain many balls of pure native iron, which is also commonly found in iron samples.

All this data on the iron formations is taken from [Dmitriev, 2016]. These formations are not dated (time-stamped). However Leonid Kulik wrote [Kulik, 1939], translated by A.O.:

"...; immediately after the meteorite fall, the Evenks found pieces (the size of a fist) of white, shiny metal on the charred soil in the center of the forestfall;..."

Interestingly that Kulik did not present details of these accounts. Anyway, all hopes are for future expeditions.

**Microparticles.** There were many works on microparticles in Tunguska. Only some of the works are considered here which devoted to microparticles outside of the forestfall region. In 1961-1962, KMET collected a large number of soil samples at a distance of up to ~200 km from the epicenter, the samples were processed at a special enrichment equipment and the number of microspherules was calculated in the enriched fraction. KMET concluded that at a distance of ~70 km from the epicenter, an area of increased counting concentrations of the microspherules begins in a northwesterly direction, which extends further in a northwesterly direction by at least 200 km, and maybe even further. According to KMET, it was dust of the Tunguska comet [Fesenkov, 1968].

However, the samples collected by KMET were not time-stamped, meaning that they could have accumulated microspherules over centuries, if not longer. Therefore, in the mid-1960s, KSE switched to studying peatlands using deep cuts, which allowed time-stamping.

Here is what Igor Doroshin wrote about this research by KMET [Doroshin, 2012], translated by A.O. (TM is Tunguska meteorite):

"Criticism of this work is given in [Ivanova, et al. 1967]. It turned out that the maximum counting concentrations of the balls/microspherules in this work correspond to the usual counting concentrations of other areas of the Earth, i.e., represent an ordinary background. It was also proven that the values of count concentrations in "empty" samples are far from the true values, which was established by direct verification of previously processed samples. It is possible to add and that argument, that no falling down in sizes along the whole trail was observed, otherwise this circumstance would be necessarily marked by researchers as evidence of the involvement of the found balls/microspherules to TM. Thus, both the structure itself and relationship of the found particles to TM are questionable. In addition, in 1986, peat samples were collected on a profile that crosses the specified structure near the mouth of the Korda-Mutorai River. None of the 4 samples taken from the supposed trail did not show any excess of magnetite balls/microspherules over the background."

KSE also conducted search for microspherules, several types of them were discovered, but none of them were declared as remnants of the alleged Tunguska spacebody. It is noteworthy that gas bubbles were found in some of the microspherules, and the composition of the gases resembles that of the gases in tektites, particularly hydrogen. The Author has considered the possible role of endogenic hydrogen degassing in Tunguska [Ol'khovatov, 2025e].

In [Fiałkiewicz-Kozieł et al., 2016] results were presented about research of dust in the Mukhrino bog located ~60.9° N, ~68.7° E, i.e. about 1800 km to the west from the epicenter. Here are some fragments from [Fiałkiewicz-Kozieł et al., 2016] (CAR is carbon accumulation rates):

"The time of abrupt change in CAR (from 22.3–244  g m− 2 y− 1), dust flux and several other proxies are noted in the modelled age of the 57–58 cm layer (cal AD 1882 ± 43 – 1920 ± 28) and is in good agreement with the date of the Tunguska cosmic body event (TCB) which happened in June 1908 (Fig.2)... .

Indirect evidence of TCB-induced dust fallout in the Mukhrino peatland is an unusual occurrence of mullite in the "Tunguska layer" and its absence in adjacent layers (Supplementary Table 2). Mullite is a high-temperature phase that forms due to the decomposition and transformation of clay minerals at temperatures > 1100 °C [40] . The mineral, together with microspherules, scoria like objects (SLOs) and other high temperature minerals (corundum, suessite) lacking in the 'Tunguska layer', has been deemed evidence of a Younger Dryas meteorite impact[41]. In the 'Tunguska layer', mullite may be a product of melting of dust and soil minerals at extremely high temperatures induced by the TCB explosion. A distinct peak in microscopic charcoal contents (Fig. 4), indicative of distant fires, has been recorded in this layer together with an increased concentration of Se (Supplementary Table 1), a biogenic element released during forest fires[42]. It is known that the TCB impact set 2000 km$^2$ of taiga on fire. The main stream of dust formed during the TCB explosion passed westwards through Siberia, Europe and America[7]. As to the best of our knowledge, there were no local fires during this time in the vicinity of the Mukhrino peat-land, it is proposed that post TCB fires could have influenced the dust flux and the element concentration in this peat layer. The layer is characterised by the

highest Th/U value (3.9; Supplementary Table 1) which would indicate a change in the supply of natural dust."

There is some exaggeration in the words "set 2000 km$^2$ of taiga on fire" - the forest fire in Tunguska event was rather moderate with possible exception in  some places near the epicenter. Some other interpretations in the fragment are problematic in the Author opinion.

Anyway, the appearance of mullite is very interesting, because in nature mullite is met rather rare - usually in some geological structures and also produced by lightning strikes in sand.

**Forestfalls.**  The Author described forest-falls in Tunguska in [Ol'khovatov, 2021; 2025c]. Here the Author would like to supply some additional info.

Yu. D. Lavbin conducted expeditions in the mid 1990s in the Kazhma river region (~59.6˚ N; ~96.1˚ E), where satellite images revealed alleged forestfall. Lavbin delivered his report at the 1998 conference on the 1908 Tunguska event in Krasnoyarsk ( http://tunguska.tsc.ru/ru/science/conf/1998/6/9/ ). Here is a fragment of the report translated by A.O. (for more translated fragments see [Ol'khovatov, 2021]):

"During field work in the area of the epicenter and not far from it, trees were cut down, which showed that a space disaster in this area occurred in 1908,... <...>

The analysis of the soil, samples of which were taken in this area, in different places, as well as the spherules found in it, showed the anomalous presence of many elements, some of which are truly cosmic. In particular, the percentage of Iridium has a value 3-4 orders of magnitude higher than its clark content in soils and rocks of the Earth. At the same time, a high content of such elements as: Germanium, Indium, Cobalt, Boron, Barium, Molybdenum, Manganese, Nickel, Lead, Copper, Magnesium, Zinc, Titanium, Sodium, Calcium, Phosphorus, etc., Iron 50 % (total) was found (Table 1)."

The Author remembers the following.  It happened in 1998 in Krasnoyarsk during the conference dedicated to the 90th anniversary of Tunguska. That's why the Author doesn't remember it word for word, but remembers the gist of it (that's why the direct speech below is approximate). Probably during the conference-break the Author saw how a small group of Tunguska researchers led by Nikolai Vasil'ev is discussing something. The Author approached - it turned out that Vasil'ev was holding a sheet of paper with the Lavbin's data on the iridium content in the site of the forestfall, which, according to Lavbin, was the site of the 1908 falls. Vasil'ev was surprised at the high concentration of iridium in this place. Someone (possibly E.M. Kolesnikov, but the Author doesn't remember exactly) said something like this: "But this is Lavbin!" To this Vasil'ev objected: "But he measured it in laboratories" and, for some reason, looked at the Author. The Author said nothing...

Indeed Lavbin (as a scientific researcher) was rather controversial person. He was a businessman who was interested in Tunguska, and made some statements about an alien's spacecraft, etc.

In his another report to the 1998 conference in Krasnoyarsk on Tunguska (can be read at https://tunguska.tsc.ru/ru/science/conf/1998/6/8/ ) Lavbin wrote the following (translated by A.O.):

"When conducting radiometric surveys on the ground, the radiation situation in the area, especially in areas where the earth's surface has been damaged, is quite strange and intriguing. Away from bodies of water or marshy depressions, the background radiation level is 10 to 15 microroentgens/hour. As one approaches these formations, the radiation level decreases, and along their shores and terrain, it is 1-3 microroentgens/hour. Further research has shown that elevated levels of the element beryllium were detected in these areas, but whether this was the cause of the change in radiation remains to be determined.

Magnetometric data in this area exhibit superanomalous features. A near-giant magnetization of local rocks (traps) was detected, as well as their reversal of magnetization. According to expedition member Eduard Nikolaevich Lind, kandidat nauk (~PhD) in Geology and Mineralogy and head of the petrographic laboratory at KNIIGIMS, this anomaly can only be compared to the impact of lightning. However, lightning produces a pinpoint effect, and in this area, this anomaly is practically ubiquitous.

The enormous magnetization of rocks and soils in this area, as well as their reversal of magnetization, most likely occurred as a result of the electromagnetic impact of a cosmic object on the Earth's surface. The level of rock magnetization in some places reaches 65-70 nT. Therefore, it can be confidently stated that this electromagnetic impact of the object caused a geomagnetic storm in Eastern Siberia on June 30, 1908."

E.N. Lind wrote in his report on the 1998 Krasnoyarsk conference on Tunguska (can be read here: https://tunguska.tsc.ru/ru/science/conf/1998/5/4/ ), translated by A.O.) :

"According to this method, although using a proton and quantum magnetometer, large blocks of intrusive traps with a chaotic distribution of residual magnetization vectors (even within a single block) have been found in the north of the Boguchansky District along the supposed route of the Tunguska body, on a number of local sites along the shores of lakes of unclear genesis. Moreover, the intensity of magnetization is very high at low values of magnetic susceptibility, which is typical for isothermal magnetization. Usually, such reversal of magnetization of rocks are found only in watershed areas."

Some years ago, the Author tried to find out the details of dating the wood and called E.N. Lind. Unfortunately, many years after the expedition, he could not remember exactly and suggested that dating was probably in the Forest Institute in Krasnoyarsk (nowadays: Federal Research Center "Krasnoyarsk Science Center of the Siberian Branch of the Russian Academy of Sciences").

In 1948 an article was published by P.L. Dravert about so called the Ketskii forestfalls. Here the Author closely follows the description from [Vasil'ev et al., 1981]. The forestfalls were described by P.L. Dravert, based on the words of the soil scientist M.A. Sergeev, who traveled along the Ket River (in the northern part of the Tomsk Region) with a land surveying expedition in 1932. The situation was as follows. Sergeev traveled a significant distance along the banks of the Ket River, between the village of Maksimkin Yar and the area of the former Ob-Yenisei channel. During the passage route, near the village of Maksimkin Yar, Sergeev came across a huge windfall, about which Sergeev's guide hant Mizurkin said that it was caused twenty years ago terrible storm that played out in these places. According to Mizurkin, similar, but even larger-scale windfalls take place in the area of the villages of Lukyanovo and Orlyukovo on the Ket River (above villages Ozyornoe, located at the mouth of the Ob-Yenisei Canal). P.L. Dravert, based on the Sergeev's story, suggested that the cause of the Ketskii windfalls was the Tunguska meteorite, which fell in the form of several boulders in a vast area of Central and Western Siberia.

KSE was interested to investigate the alleged forestfalls. In the late 1950s G.F. Plekhanov said [Erokhovets, 1960] (translated by A.O.):

"In the upper reaches of the river Ket, north of Tomsk, there is an area of the forestfall. Moreover, as they say, this forestfall is similar to the destruction in the Tunguska catastrophe area. Some windfalls relating to about 1908 are on the river Korda. They are described by Dravert ... So, out of curiosity, I took a globe and connected these points from the river Ket to the place of falling the Tunguska meteorite. It turned out a rather straight line. What is it - randomicity? May be. But the randomicity is strange and interesting. Need to check? Necessarily."

The Ket river basin was investigated in 1960 by an expedition [Vasil'ev, et al.,1963]. Several forestfalls were discovered (dimensions can be as: length ~40 km and more, and ~4 km in width). The forestfalls were of strip-like character. Vasil'ev with colleagues concludes [Vasil'ev, et al., 1963] (translated by A.O.):

"As for the causes of the windfall in the Ket River basin, then, judging by old-timers, it is associated with two hurricanes, the first of which was between 1906 and 1912, and the second - between 1921 and 1930.

<...> Much later on the site of the windfall there was a forestfire, which is connected with those traces of burn on trees that are visible on the fallen trunks even nowadays. Thus, they are not related directly to the event, which resulted in the windfall.

Considering the fact that the direction of the tree-falls coincides with the prevailing direction of the winds in this area, it can be assumed that the specified windfall has nothing to do with the flight or fall of the Tunguska meteorite. In any case, the strip-like fall of trees is typical for windfalls, the causes of which are strong windstorms."

Interestingly, that at least one forestfall in that region occurred between 1906 and 1912 (which includes 1908...), and also that direction of the forestfall is "looking" to the side of the Kulikovskii forestfall, which is situated about 900 km away to the east. Randomicity? May be...

**Earthquakes**. In [Ol'khovatov, 2022] there were some examples of rather strong seismic phenomena at relatively large distances from the epicenter. Here the Author would like to add one more. In [Astapovich, 1948] I.S. Astapovich presented some data which he got from P. L. Dravert before a death of Dravert. Dravert wrote about an earthquake on June 30, 1908. It should be noted that the number of opening quotation marks in [Astapovich, 1948] exceeds the number of closing quotation marks, and it is necessary to guess which text is being referred to. However, since the focus is on the facts, here are a few excerpts from [Astapovich, 1948], translated by A.O.:

"In the summer of 1915, having arrived in Yeniseysk to prepare for a trip to the Pito-Angarsky region on a business trip of the AN {probably Academy of Science - A.O.}, I met the former gold-miner N.E. Matonin. From his stories about natural phenomena that he had witnessed, the story about the earthquake in the South Yenisei taiga especially stopped my attention. I pass on what has survived in my memory.

According to Matonin, he lived in 1908 at his gold mine on the Kadra River, which flows into Bolshoi Pit (the right tributary of the Yenisei River). One morning, while still in bed, he was suddenly thrown up by a strong jolt, "underground" blows were heard, the walls of the house shook, the glass in the frames rattled. Frightened, Matonin rushed out of the house, deciding that it was necessary to escape from the earthquake by the water. When he reached the pond, he saw that the surface of the water was moving like waves in the wind. He lost his composure. After a while, the earthquake stopped. As it turned out, some items in the cupboard had fallen, a picture in a frame had tilted, and a clock with an external pendulum had stopped. The pets, like the people, were in a state of great confusion. In the surrounding area, some rocks had collapsed from the cliffs. The Tunguses, who came from the east every summer to purchase supplies, had reported terrifying phenomena in the taiga, including an earthquake that had caused widespread fear.

"Later, while traveling through the South Yenisei taiga for mineralogical purposes, I visited a number of mines and settlements along the Kadra, Ayakhta, Gorbilok, Pechenga {probably Penchenga -A.O.}, Uderei, and their tributaries: Tatarka, Bolshaya Murozhnaya, and several others. Along the way, I asked about the earthquake. In one location, I was able to determine the date of the earthquake from a notebook: June 17, Old Style. The people I interviewed were diverse, including a doctor, two engineers, and a mining inspector. I don't remember the details, but I can tell you that the earthquake was felt throughout the South Yenisei taiga, with varying intensities, numbers of tremors, and durations. The area affected by the earthquake lies approximately between 58° 10' and 59° 20' N and 93° 00' and 94° 45' E, and possibly further east. The earthquake reached a Rossi-Forel intensity of VI. <...> According to Matonin, this earthquake also occurred in the North Yenisei taiga: I remember the rivers Chirimba and Enashimo (right tributaries of the Bolshoi Pit)."

There are also reports about earthquakes in the mines in the South Yenisei taiga which were published in 1908 in a local newspaper [Ol'khovatov, 2022]. The mine buildings made a cracking and creaking sound, so that people ran out into the street with fear. Workers who were at work noticed how the kulibinki (gold washing machines) were shaking and dust rose from the ground, which caused panic and runaway from work. At the Gavrilovskii mine, horses fell to their knees.  Dishes were falling from the shelves on the Zolotoi bugorok mine. The Gavrilovskii mine was about 493 km from the epicenter at azimuth 266° , the Zolotoi bugorok mine was about 493 km from the epicenter at azimuth 267°.

The observer  (T. Grechin) of the Shamanskie water-measuring posts wrote on June 18 i.e. on the next day after the event (translated by A.O.):

"On June 17, at 8 o'clock in the morning, there were some strong blows several times in the north side like thunder, from which the windows trembled in the frames, the trees bent down and the leaves of them shook; at the same time it was clear and quiet, and even the water did not lose its gloss, no damage was noticeable. As local peasants who were at field work at that time told me, they saw some kind of a fireball flying in the north side, from which such strong blows like explosions seemed to occur."

The words "The trees bent down" (and without any noticeable air-action) are a very remarkable fact, as well as the fireball. The Shamanskie water-measuring posts are situated about 586 km at azimuth about 180° from the epicenter.

A priest D.A. Kazanskii from the settlement Zhimygytskii stan (1010 km from the epicenter at azimuth 176°) reported that from 8 a.m. to 8 h. 20 min. a.m. majority of people felt a slight shivering. There were like rolls of thunder, and then cannon's shots far away. He evaluated the earthquake by Rossi-Forel scale as 3.

There was even some minor damage at places the "Tolstyi mys" (~1023 km from the epicenter at azimuth 169°) and the "Troinaya guba" (~1019 km from the epicenter at azimuth 170°), which are even farther than Irkutsk relative to the epicenter [Ol'khovatov, 2022].

The above facts allow to conclude that the source of the seismic waves (having a magnitude of 4.5-5.0) in the epicenter region (caused by the destruction of a hypothetical Tunguska meteoroid in the air) could not have produced the observed seismic phenomena. Indeed:

a) as shown in [Ol'khovatov, 2022], the amplitude of the vibrations in the areas under consideration is too large for such a distance from the epicenter of such a seismic source. In addition, the large region affected by the tremors makes it highly unlikely that the large amplitude is caused by the geological structure of the specific area.

b) This applies not only to the amplitude, but also to the frequency of seismic vibrations, which is significantly higher than for seismic waves propagating from the epicenter of such a seismic source [Ol'khovatov, 2022].

c) The most remarkable thing is the time of occurrence of these seismic events, which range from 23.43 GMT (June 29) to approximately noon - see [Ol'khovatov, 2022]. While it is possible to assume a time error of ~30min for 23.43 GMT, it is not possible for noon.

By the way, here are other accounts of the earlier events. Here is one of them. The event took place in the town of Kirensk ( 492 km from the epicenter at azimuth 132° ) on the Lena river. Here is how Bronshten describes the event [Bronshten, 1997] (translated by A.O.):

"Ivan liked to get up early and do jogs in one verst. June 30, 1908 morning was not an exception. This morning was cloudless, the sun brightly shone, no any wind. Suddenly Ivan's attention was drawn by the amplifying noise proceeding as it seemed to him, from southeast side of the sky. Neither from the East, nor from the North, nor from the West nothing similar was felt. The sound came nearer. "All this began." - Ivan Suvorov wrote, - "on my watches verified the day before by post-office of Kirensk, at 6 hours 58 minutes local time. Gradually coming source of noise began to be listened from South-South-West side and passed into the West-North-western direction that coincided with the shot-up fiery column up at 7 hours 15 minutes in the morning". <…> What surprises us in these accounts? First of all, time of the beginning of audibility of an abnormal sound - 6 hours 58 minutes while the fiery column shot up, in full consent with other definitions, at 7 hours 15 minutes. The Tunguska bolide could not fly, making a sound, for 17 minutes. During this time at a speed of 30 km/s it would fly by 30000 km, that is at 6 hours 58 minutes it was far outside the atmosphere and could not make any sounds. It means, this moment belongs not to the beginning of emergence of a sound, and to some other event, for example to Ivan's exit from the house.

The correct indication of the moment of explosion forces us to reject also all other possible assumptions: for example, that watches of Ivan lagged behind per day for 17 minutes or that local time of Kirensk strongly differed from local times of other points. Moreover, - in the same Kirensk the director of a meteorological station G. K. Kulesh recorded according to indications of a barograph arrival of an air wave (i.e. the same sounds) after 7 hours.

So inexact Ivan defined and the direction from where the sounds came. The Tunguska bolide flew by, by the most exact definitions, to the North from Kirensk. The closest point of a trajectory was from it to the northeast. Then the bolide moved to the North and, at last, to the northwest.

According to E. L. Krinov in his book "Tunguska Meteorite" (Moscow: USSR Academy of Sciences, 1949, p. 54) many eyewitnesses later claimed that they heard the sound before they saw the bolide (which in fact could not be). Apparently, this is some kind of property of inexperienced observers who reported what they saw much later, several years after the event."

The Bronsten's attempts to "improve" the account will be not commented, just adding that the barograph in Kirensk recorded arrival of the air-pressure disturbance at 7 h. 48 m., and the barograph detects air-pressure disturbances of very low frequency.

The above facts point to a swarm of earthquakes. Some of them were accompanied with earthquake lights. It is noteworthy that this point of view was widespread in the region. For example, engineer V.P. Gundobin reported in 1924 that in 1921 Ivan Andreevich Kochergin (a resident of the village of Shchekino) claimed that he witnessed an "earthquake" accompanied by a rumble and light phenomena [Vasil'ev et al., 1981].

At least one of the earthquakes was recorded  by the Irkutsk seismic station. On the recording of the Irkutsk seismic station I.P. Pasechnik marked pos. "P" as arrival of longitudinal waves of a local earthquake. This took place about 2 hours later after arrival of the "Tunguska earthquake" waves [Ol'khovatov, 2022].

Peculiarities of the global seismicity were considered in [Ol'khovatov, 2025e].

There were also some meteorological peculiarities in the region. Some of them already were considered by the Author in previous works - see, for example, [Ol'khovatov, 2021; 2023b]. Remarkably, for example, a weather log of the Kirensk meteostation (situated about 491 km to the SE from the Tunguska epicenter ) marked in the afternoon of June 30 a strong wind. For comparison, the previously strong wind was noted by this meteorological station in 1908 only on March 31, and next time - December 9th. So June 30, 1908 was a peculiar day in the region indeed.

Analysis of the meteorological peculiarities is beyond the scope of this paper. Here the Author just wants to point to some analysis by other researchers. Statistical analysis of about 700 accounts (collected in various years) conducted in [Demin et al., 1984] revealed that (translated by A.O.):

"The Tunguska phenomenon, according to eyewitnesses, was accompanied by various meteorological phenomena (Table 7). The most frequently observed were "strong wind", as well as "haze, fog, fog". The complexity of atmospheric processes is evidenced by sharp changes in air temperature recorded in a number of accounts. Thunderstorms, individual lightning discharges, local development of windstorms, hurricanes and whirlwinds are noted."

## 3. Discussion

Various phenomena were not limited to the vicinity of the epicenter, but occurred over a large region and for at least several hours. Moreover one of the "later" earthquakes was even recorded by the Irkutsk seismic station.

It would be important to research the Kazhma river forestfall. If the information reported by Lavbin is confirmed, it will shed new light on the Tunguska event.

The finds of various substances are of particular interest. Unfortunately, there is no dating of the appearance of the iron formations. The Kulik's reference to Evenk's words gives only a hint that the appearance coincided with the Tunguska event. The native iron formations were discovered in the Siberian traps [Tomshin et al., 2023]. The reason for the appearance of native iron formations on the planet's surface is still being discussed [Tomshin et al., 2023]. The Author would like to draw attention to similar phenomena in miniature [Ol'khovatov, 2020b]. Anyway it looks like something really intriguing happened in Tunguska.

As for mullite, it is dated to 1908 in [Fiałkiewicz-Kozieł et al., 2016]. Regarding the mullite interpretation in [Fiałkiewicz-Kozieł et al., 2016] the Author is a bit cautious, due to several points, including this one from [Fiałkiewicz-Kozieł et al., 2016]:

"...dust flux and several other proxies are noted in the modelled age of the 57–58 cm layer (cal AD 1882 ± 43 – 1920 ± 28) and is in good agreement with the date of the Tunguska cosmic body event (TCB) which happened in June 1908..."

Therefore, the year 1908 may not be included in the specified depth interval due to the uncertainty in the dating. Let's assume, given the uniqueness of both the Tunguska event and the mullite discovery in the peat, that the dating corresponds to the expected one. What could be the cause of the mullite appearance? According to [Schneider et al., 2008]:

"Due to its high temperature but low pressure formation conditions, mullite occurs very rarely in nature. It has been found at the contact of superheated magma intrusions with $Al_2O_3$-rich sediments, as on the Island of Mull (Scotland), where the name mullite comes from. Mullite has also been described in high temperature metamorphosed rocks of the sanidinite facies[1] and in hornfelses (porcellanite), e.g., at the contact of bauxites with olivine dolerite intrusions. Special and rare occurrences of mullite are in alumino silicate lechatelerite glasses produced by lightening impact in sandstones[2], and in small druses of volcanic rocks (e.g., in the Eifel mountain, Western Germany), where it probably grew under moderate hydrothermal conditions...."

The Author is not aware of any publication reporting about mullite near the epicenter, but this may be a matter of terminology. Anyway, some particles exposed to high temperatures were discovered [Kletetschka et al., 2025].

Let's try to consider how mullite could appear near the epicenter. The forest-fire in Tunguska was not intensive with exclusion several places near the epicenter. Moreover there were much more powerful forest-fires (in various years) in taiga even closer to the Mukhrino bog, and they did not deposit mullite in the bog.

Anyway mullite was discovered in volcanic rocks ~350 km westwards from the epicenter [Sokol et al., 2019]. Therefore, it is reasonable to check if it was already near the epicenter long before the event.

Electric discharges [Ol'khovatov, 2025g] could be another source of mullite.

The third way could be formation due to hydrogen (and possibly other gases) outbursts burning [Ol'khovatov, 2025g]. Indeed even oil–gas fire can result in temperatures no lower than 1200 ºC [Kokh et al., 2016].

In any case, due to the large amount of time uncertainty in the dating, which includes many events, such as the massive fire in the taiga in 1915, it is difficult to draw any definitive conclusions about the mullite.

## 4. Conclusion

Various manifestations of the Tunguska event were not limited to the vicinity of the epicenter, but occurred over a large region and for at least several hours.

The general conclusion is that the Tunguska event was a very complex phenomenon. Research of the Tunguska event requires the participation of experts in various fields. In the opinion of the Author, researching the Tunguska event will allow us to better understand the life of such a complex system as our planet.

## ACKNOWLEDGEMENTS

The Author wants to thank the many people who helped him to work on this paper, and special gratitude to his mother - Ol'khovatova Olga Leonidovna (unfortunately she didn't live long enough to see this paper published...), without her moral and other diverse support this paper would hardly have been written.